# Force unfolding kinetics of RNA using optical tweezers. II. Modeling experiments.

M. Manosas, J.-D. Wen, P. T. X. Li, S. B. Smith, C. Bustamante, I. Tinoco, Jr., F. Ritort

## Abstract

By exerting mechanical force it is possible to unfold/refold RNA molecules one at a time. In a small range of forces, an RNA molecule can hop between the folded and the unfolded state with force-dependent kinetic rates. Here, we introduce a mesoscopic model to analyze the hopping kinetics of RNA hairpins in an optical tweezers setup. The model includes different elements of the experimental setup (beads, handles and RNA sequence) and limitations of the instrument (time lag of the force-feedback mechanism and finite bandwidth of data acquisition). We investigated the influence of the instrument on the measured hopping rates. Results from the model are in good agreement with the experiments reported in the companion article (1). The comparison between theory and experiments allowed us to infer the values of the intrinsic molecular rates of the RNA hairpin alone and to search for the optimal experimental conditions to do the measurements. We conclude that the longest handles and softest traps that allow detection of the folding/unfolding signal (handles about 5-10 Kbp and traps about 0.03 pN/nm) represent the best conditions to obtain the intrinsic molecular rates. The methodology and rationale presented here can be applied to other experimental setups and other molecules.

**ABBREVIATIONS**
F, folded; U, unfolded; bp, base-pair; F-U, folding/unfolding; ssRNA single-stranded RNA; PM, passive mode; CFM, constant-force mode; WLC, worm-like-chain; IFE, ideal-force ensemble; ME, mixed ensemble; SNR, signal-to-noise ratio.

# 1. Introduction

Recently developed single-molecule techniques (2) have been used to exert force on individual molecules, such as nucleic acids (3,4,5,6,7) and proteins (8,9,10). These techniques make it possible to test the mechanical response of biomolecules which can be used to obtain information about their structure and stability. Moreover, the study of the kinetics, pathways, and mechanisms of biochemical reactions is particularly suited to single-molecule methods where individual molecular trajectories can be followed (11,12,13).

Optical tweezers have been used to study folding/unfolding (F-U) of RNA hairpins (14,15,16,17). The experimental setup consists of the RNA molecule flanked by double-stranded DNA/RNA handles; the entire molecule is tethered between two polystyrene beads via affinity interactions. The handles are polymer spacers required to screen interactions between the RNA molecule and the beads and to prevent direct contact of the beads. One of the beads is held in the optical trap; the other bead is controlled by a piezoelectric actuator to apply mechanical force to the ends of the RNA molecule. In hopping experiments a given constraint, i.e. a fixed force or a fixed extension, is applied to the experimental system while both the force and the extension of the molecule are monitored as a function of time. Close to the transition force (around 10-20 pN for RNA or DNA hairpins at room temperature (13,14)), a hairpin molecule can transit between the folded (F) and the unfolded (U) states, as indicated by the change in the molecular extension: the longer extension represents the unfolded single-stranded conformation; the shorter one to the folded hairpin. From the lifetimes of the single RNA molecule in each of the two states, we can obtain the rates of the F-U reaction (1,14). Both the unfolding and folding rate constants are force-dependent following the Kramers-Bell theory (17,18,19). From their ratio the force dependent equilibrium constant for the F-U reaction can be obtained.

In order to obtain accurate information about the molecule under study it is important to understand the influence of the experimental setup, including the handles and the trapped bead, on the measurements. In a recent simulation, Hyeon and Thirumalai (20) examined the relationship between the amplitude of the F-U transition signal and the magnitude of its fluctuations at various handle lengths. On the other hand, experimental results have shown that the F-U kinetics was dependent on the trap stiffness (21). Several questions then arise: how different is the measured rate from the intrinsic molecular rate, *i.e.*, the F-U rate of the RNA in the absence of handles and beads? What are the optimal working conditions to obtain the intrinsic molecular rates? To address such questions, we previously proposed a model (22), which considered the effect of the trapped bead and the handles on a two-state RNA folding mechanism. In this work, we further advance our simulation by incorporating a mesoscopic model introduced by Cocco et al. (23) that takes into account the sequence-dependent folding energy. We have then applied this model to a simple hairpin, P5ab (14). We investigate how the measured rates vary with the characteristics of the experimental setup and how much they differ from the intrinsic molecular rates of the individual RNA molecule. In a companion paper, we have also measured the F-U kinetics of the RNA hairpin by optical tweezers (1). The theoretical and experimental results agree well.

The organization of the paper is as follows. In Sec. 2 we introduce the model for the experimental setup and describe its thermodynamic properties. We also analyze the characteristic timescales of the system. In Sec. 3 we discuss the influence of the different elements of the experimental setup on the kinetic rates. Limitations of the instrument which affect the measured F-U rates, such as the force-feedback time lag and the data acquisition bandwidth, are also considered. Based on the various timescales of the different dynamical processes described in Sec. 2, we develop a kinetic model for the RNA hairpin and a numerical algorithm used to simulate the hopping dynamics in Sec. 4. In Sec. 5 we carry out a detailed analysis of the dependence of the kinetic rates on the characteristics of the experiment (such as the length of the handles and the stiffness of the trap), and compare our simulation results with the experimentally measured F-U rates. A search of the best fit between theory and experiments allows us to predict the value of the intrinsic molecular F-U rate of the RNA molecule. Finally, we discuss what are the optimal experimental conditions to minimize the effect of the instrument and to obtain the intrinsic molecular rates.

## 2. Experimental setup and experimental modes

Hopping experiments (1) were done with a single RNA hairpin P5ab, a derivative of the L-21 *Tetrahymena ribozyme*. The kinetics of this RNA with 1.1 Kbp handles had been studied previously (14). In Fig. 1 we show a schematic picture of the setup used in such experiments. To manipulate the RNA molecule two RNA/DNA hybrid handles are attached to its 5'- and 3'-ends. The free ends of the handles are attached to micron-sized polystyrene beads. One bead is held fixed in the tip of a micropipette while the other bead is trapped in the focus of the laser, which is well described by a harmonic potential of stiffness $\varepsilon_b$. The configurational variables of the system $x_b$, $x_r$, $x_{h_1}$ and $x_{h_2}$ are the extensions of each element (trapped bead, RNA molecule and handles respectively) along the reaction coordinate axis (i.e. the axis along which the force is applied). The external control parameter $X_T$ is the distance between the center of the optical trap and the tip of the micropipette. In an experiment, the applied force $f$ and the distance $x_b$ are measured. From the value of $x_b$ the changes in the distance between the two beads $x$, corresponding to the end-to-end distance of the molecular construct formed by the two handles and the RNA molecule (Fig. 1), can be obtained; $x = X_T - x_b - R_{b1} - R_{b2}$, where $R_{b1}$ and $R_{b2}$ are the beads radii. A more detailed description of the experimental setup is given in (1).

In hopping experiments, the force $f$ and the changes in the extension $x$ as a function of time are recorded. The structural changes of the RNA molecule can be identified with the sudden changes in force and extension, here referred to as $\Delta f$ and $\Delta x$ respectively.

Experiments are carried out in two different modes: the passive and constant-force modes. In the passive mode (PM) the distance $X_T$ between the center of the trap and the tip of the micropipette is held fixed. In PM hopping experiments both the extension $x$ and the force $f$ hop when the molecule switches from one state (F or U) to the other. In the constant-force mode (CFM) the force is maintained constant by implementing a force-feedback mechanism. In CFM hopping experiments the changes in the state of the RNA molecule can be identified with the measured changes in the extension $x$ of the molecular construct. Experimentally, P5ab folds and unfolds with no apparent intermediates (1,14). The experimental traces show jumps in force and extension, $\Delta f$ and $\Delta x$, that correspond to the full unfolding or folding of the RNA hairpin. From the data we can extract the mean lifetimes of the F and the U states of the molecule, $\tau_F$ and $\tau_U$, at a given force. The folding and unfolding rates, $k_F$ and $k_U$, are the reciprocal of $\tau_U$ and $\tau_F$, respectively.

## 2.1 Thermodynamics of the experimental system.

The experimental setup is modeled as previously described (22). The bead confined in the optical trap is considered as a bead attached to a spring whose stiffness equals the trap stiffness, $\varepsilon_b$, and the double-stranded DNA/RNA handles are modeled by the worm-like-chain (WLC) theory (24, 25), which describes the elastic behavior of polymers by two characteristic parameters: the contour (L) and the persistence (P) length. In our previous model (22), we considered a two-state model for the F-U of an RNA molecule. Here we extend that approach by including intermediate configurations of the hairpin where a partial number of bps are opened sequentially starting from the end of the helix. In this description, the molecule can only occupy intermediate configurations in which the first $n$ bps are unpaired and the last $N-n$ are paired, where $N$ is the total number of bps in the native hairpin. The index $n$ is used to denote such intermediate configurations (Fig. 2), e.g. the F state corresponds to $n = 0$ and the U state to $n = N$. This representation excludes the existence of other non-sequential *breathing* intermediate configurations that might be relevant for thermal denaturation (26). For a given value of the control parameter (generically denoted by $y$, e.g. $X_T$ or $f$), and for each configuration $n$ of the RNA molecule, we can define the thermodynamic potential $G(y,n)$ as (22):

$$G(y,n) = G^0(n) + G'_y(n) \qquad (2.1)$$

where $G^0(n)$ is the free energy of the RNA hairpin at the configuration $n$ and $G'_y(n)$ describes the energetic dependence of the experimental system on the control parameter. Note that the term $G'_y(n)$ is sequence independent, so all information about the sequence is included in the term $G^0(n)$. The critical control parameter ($F^c$ or $X_T^c$) is the value of the control parameter at which the F and U states are equally populated. For the P5ab hairpin, the value of the critical force measured in the experiments is around 14.5 pN (1).

In the ideal-force ensemble (hereafter referred as IFE), the force exerted upon an RNA hairpin is the control parameter ($y = f$) and the system reduces to the naked RNA molecule without beads and handles. The contribution $G'$ to the free energy of the RNA molecule is given by (27):

$$G'_f(n) = W_r(x_r(f,n)) - fx_r(f,n), \qquad (2.2)$$

where $W_r(z)$ is the work required to stretch the molecular extension $x_r$ of the ssRNA from $x_r = 0$ to $x_r = z$. In our experiments (Fig. 1), where handles, beads and the RNA molecule are linked, the natural control parameter in Eq 2.1 is $y = X_T$. This defines what has been denoted as the mixed ensemble (hereafter referred as ME) (28). In such case the contribution $G'$ in Eq. 2.2 has been derived in (22):

$$G'_{X_T}(n) = V_b(x_b(X_T,n)) + \sum_{i=1,2} W_{h_i}(x_{h_i}(X_T,n)) + W_r(x_r(X_T,n)), \qquad (2.3)$$

where $x_\alpha(X_T,n)$, with $\alpha = b,h_1,h_2,r$, is the mean value of $x_\alpha$ for a given value of the control parameter $X_T$ and for a given configuration $n$ of the RNA hairpin. $V_b$ represents the optical trap potential, $V_b = \frac{1}{2}\varepsilon_b x_b^2$ and $W_\alpha(z)$, with $\alpha = h_1,h_2,r$, is the work done upon each of the handles and the ssRNA to stretch their molecular extensions from $x_\alpha = 0$ to $x_\alpha = z$,

$$W_\alpha(z) = \int_0^z f_\alpha(x)dx, \qquad (2.4)$$

where $f_\alpha(x)$ is the equilibrium force extension curve for the element $\alpha$ (22). These different contributions to the thermodynamic potential are free-energies corresponding to the trapped bead, the handles and the ssRNA molecule. Therefore, in the ME, the thermodynamic potential given by Eq. 2.1 depends not only on the RNA properties but also on the characteristics of the different elements of the setup, such as the stiffness of the trap and the contour and persistence lengths of the handles. In order to extract thermodynamic information of the RNA molecule from the experimental results, we need to take into account the contribution from each of the elements forming the setup (22). In the Supplementary Materials we show how the shape of the thermodynamic potential (Eq. 2.1) is modified for different values of the stiffness of the trap and the length of the handles. The characteristics of the experimental setup change the value of the maximum of the free-energy along the reaction coordinate, which is related to the kinetic barrier separating the F and U states, and thus influences the kinetics of the F-U reaction. The dependence of the F-U rates of a DNA hairpin on the stiffness of the trap has been already reported (21).

In particular, when the experimental system gets softer, the fluctuations in force decrease and the ME approaches the IFE. The free energy landscape $G(X_T^c,n)$ converges to $G(F^c,n)$ corresponding to the IFE, in the limit where the effective stiffness $\varepsilon_{\text{eff}}$ of the whole experimental system vanishes. The effective stiffness $\varepsilon_{\text{eff}}$ is computed as:

$$\varepsilon_{\text{eff}}^{-1} = \left(\frac{d\langle f(X_T)\rangle}{dX_T}\right)^{-1} = \left[\varepsilon_b^{-1} + \varepsilon_x^{-1}\right], \tag{2.5}$$

where $\varepsilon_b$ is the stiffness of the trap and $\varepsilon_x$ is the rigidity of the molecular construct (i.e. the molecule of interest plus handles, see Fig. 1). Therefore the thermodynamics of systems with longer handles (i.e. softer handles) and softer traps approaches to the IFE case, as shown in Fig.S1 in the Supplementary Materials. However, thermodynamics alone is not sufficient to understand the influence of the experimental setup on the kinetics. For this we have to consider a kinetic description of the system. This is the subject of the next sections.

## 2.2 Timescales in the system

The dynamics of the global system presented in Fig. 1 involves processes occurring at different timescales. Therefore, in order to study the kinetics, it is essential to analyze the different characteristic times of the system: the relaxation time of the bead in the trap $\tau_b$; the relaxation time associated with the elastic longitudinal modes for the handles and the ssRNA, denoted by $\tau_{\text{handles}}$ and $\tau_{\text{ssRNA}}$ respectively; the time $k_{\text{F-U}}^{-1}$ in which the RNA hairpin folds and unfolds; the base-pair (bp) breathing time $k_{\text{bp}}^{-1}$. Table 1 reviews the different characteristic times of the experimental system.

- ***Bead:*** The time at which the bead in the optical trap relaxes to its equilibrium position is given by (22):

$$\tau_b = \frac{\gamma}{\varepsilon_x + \varepsilon_b}, \tag{2.6}$$

where $\gamma$ ($\gamma = 6\pi R_{b1}\eta$, where $\eta$ is the viscosity of water) is the frictional coefficient of the bead, $\varepsilon_b$ and $\varepsilon_x$ are the stiffness of the trap and the molecular construct respectively. Typical experimental values are: $\varepsilon_b \approx 0.02 - 0.15$ pN/nm for the trap stiffness; $R_{b1} \approx 0.5 - 1.5$ μm for the bead radius; $L_h \approx 130 - 1300$ nm and $P_h \approx 10 - 20$ nm for the contour and persistence lengths of the handles respectively, which result in values for the stiffness of the molecular construct of $\varepsilon_x \approx 0.15 - 1.5$ pN/nm (computed by using the WLC (24,25) theory at forces about 15 pN). For these values, $\tau_b$ lies in the range $10^{-5} - 10^{-3}$ s. The corner frequency of a tethered bead is defined as the reciprocal of $\tau_b$. Events that occur at frequencies higher than the corner frequency of the bead cannot be followed by the instrument.

- ***Handles and ssRNA:*** The relaxation time associated with the longitudinal modes of the handles and ssRNA when a given force $f$ is applied to their ends can be estimated from polymer theory (29) as:

$$\tau_{handles/ssRNA} = \left(\frac{k_B T}{P}\right)^{1/2} \frac{\eta L^2}{8 f^{3/2}}, \tag{2.7}$$

where $\eta$ is the viscosity of the water, ($\eta \approx 10^{-9}$ pNs/nm$^2$), $P$ is the persistence length of the double-stranded or single-stranded nucleic acids respectively, $T$ is the temperature of the bath and $k_B$ is the Boltzmann constant. For the handles used in the experiments ($P_{handles}$ = 10-20 nm) the relaxation time lies in the range $10^{-8}$ - $10^{-6}$ s. For the ssRNA corresponding to the unfolded P5ab hairpin ($N_{ssRNA}$ = 49 bases), $\tau_{ssRNA}$ is approximately $3.5 \cdot 10^{-9}$ s.

- ***RNA molecule:*** There are two different timescales associated with the kinetics of the RNA molecule. The first timescale is the overall kinetic rate $k_{F-U}$ given by:

$$k_{F-U} = \frac{k_F + k_U}{2}, \qquad (2.8)$$

where $k_F$ and $k_U$ are the folding and unfolding rates. The rate $k_{F-U}$ depends on the sequence and structural features. Under tension at which a hairpin hops, typical values of $k_{F-U}$ are in the range 0.1-100 Hz. The second timescale corresponds to the characteristic frequency for the opening/closing of single bps, $k_{bp} = \frac{k_{closing} + k_{opening}}{2}$, which is estimated to be around $10^6$ - $10^9$ Hz (30,31).

In summary, the dynamics of the system presents the following hierarchy of timescales:

$$k_{F-U}^{-1} >> \tau_b >> \tau_{handles}, \tau_{ssRNA}, k_{bp}^{-1}. \qquad (2.9)$$

Apart from the timescales associated with each of the different elements in the system there are also intrinsic characteristic timescales of the instrument. It is important to consider them in order to understand and correctly analyze the results obtained from the experiments.

- ***Instrumental times:*** There are three characteristic timescales that limit the performance of the instrument. The first timescale is defined by the bandwidth $B$ which is the rate at which data are collected in the experiments. Collected data represent an average of the instantaneous data measured over a given time window of duration $1/B$. Typical values for the bandwidth used in the experiments lie in the range from 10 to 1000 Hz. The second important characteristic timescale is given by the time lag of the feedback mechanism, $T_{lag}$, implemented in the CFM. In our experiments (1), typical values for $T_{lag}$ are 100 ms. In order to approach the IFE one would like $T_{lag}$ as small as possible. Recently, a new dumbbell dual-trap optical tweezers instrument has been developed (21,32). This design operates without feedback and can maintain the force nearly constant over distances of about 50 nm. Nevertheless, regardless of the specific instrumental design, there is a limitation in the measurement that is imposed by the corner frequency of the bead: the bead does not respond to force changes that occur faster than $\tau_b$. In our experimental setup this limiting time is approximately $10^{-4}$ s.

The third timescale ranges from seconds to minutes and corresponds to the drift of the instrument. The drift is a low frequency noise due to mechanical and acoustic vibrations, air currents, thermal expansion in response to temperature changes and other causes. Since the drift does not affect the occurrence and detection of F-U transitions, we did not take into account drift effects in our model.

## 3. Molecular and measured rates: Instrumental effects

From the force and extension traces recorded in hopping experiments (1), we can extract the rates of the F-U reaction. These traces reflect the response of the whole experimental system (Fig. 1) not only the individual RNA molecule. In addition, data collected are averaged over a bandwidth $B$, and the mechanism implemented in the CFM has a finite response time, $T_{lag}$. In this section, we analyze the effect of the experimental setup on the measured F-U rates as compared to the intrinsic molecular rates.

### 3.1 Definition of the rates

There are different experimental modes and different ways of analyzing the experimental data which result in different values of the rates of the reaction; ultimately, we wish to obtain values as close as possible to the intrinsic molecular rates. The intrinsic molecular rate, $k_{F-U}^0$, corresponds to the rate measured in an IFE where a fixed force $F^c$ (i.e., the critical force value where the F and U states are equally populated) is applied directly to the RNA molecule. In the following paragraphs, we introduce the different rates that are experimentally measurable, i.e. under the CFM and PM. These rates have been defined in our companion paper (see (1) for details).

- **CFM rates:** The CFM rates are the folding and unfolding rates measured when the instrument operates in the CFM at a given force. In what follows we will consider the critical rate $k_{CFM}^c$, which is the F-U rate (Eq. 2.8) measured at the critical force value where the molecule spends the same amount of time in the F and U states.

- **PM rates:** The force traces in the PM show that the folding and unfolding transitions occur at different forces, $f^F$ and $f^U$ respectively (Fig. 4). $f^F$ and $f^U$ are the mean forces in the upper and lower bounds of the square-like force traces respectively. The PM unfolding (folding) rate at $f^F$ ($f^U$) is then identified with the unfolding (folding) rate measured in such PM traces from the lifetime of the folded and the unfolded states respectively. The PM critical rate $k_{PM}^c$ is the F-U rate at the force value where the unfolding and folding PM rates are equal (Fig. 6).

### 3.2 Instrumental effects

To study the relation between the measured and intrinsic molecular rates, we now consider the different effects that influence the kinetics in the CFM and PM as compared with the IFE. Under the experimental conditions, the force exerted directly on the RNA molecule ($f_{RNA}$) is subject to fluctuations due to the dynamic evolution of the different elements in the experiment (Fig. 1). There are at least three contributions to these fluctuations:

*(i) Bead force fluctuations:* On timescales on the order of $\tau_b$, the force exerted upon the molecule fluctuates ($\langle \delta f_{RNA}^2 \rangle$) due to the fluctuations in the position of the bead ($\langle \delta x_b^2 \rangle$):

$$\langle \delta f_{RNA}^2 \rangle = \varepsilon_x^2 \langle \delta x_b^2 \rangle \quad , \tag{3.1}$$

where $\varepsilon_x$ is the stiffness of the molecular construct. As shown in the Supplementary Materials the effect of the fluctuations given in Eq. 3.1 is to increase the kinetics of the F-U reaction as compared with the IFE.

*(ii) Base-pair hopping effect:* At the timescale at which bps attempt to open and close, $k_{bp}^{-1}$, the bead hardly moves ($\tau_b \gg k_{bp}^{-1}$). Hence, when a bp forms (dissociates) the handles and the ssRNA stretch (contract), and correspondingly there is an increase (decrease) in the force exerted upon the RNA molecule, $f_{RNA}$. The change in the force $f_{RNA}$ after a bp opens or closes, assuming that in the timescale $k_{bp}^{-1}$ the position of the bead is fixed, is given by:

$$\sigma = \varepsilon_x \Delta x_{bp} \quad , \tag{3.2}$$

where $\Delta x_{bp}$ is the difference in extension between the formed and dissociated bp. Therefore after the formation (rupture) of a new bp the force increases (decreases) by an amount given by Eq. 3.2 and the probability to dissociate (form) it again increases as compared with the IFE case. Therefore, the *base-pair hopping effect* slows the overall F-U kinetics of the RNA molecule.

*(iii) Passive-mode force fluctuations:* In the PM the average force exerted upon the system depends on the state of the RNA molecule (Fig. 4). Therefore at the timescale $k_{F-U}^{-1}$ associated to the F-U reaction, the average force exerted on the RNA molecule will change by:

$$\Delta f_{RNA} = \Delta f = f^F - f^U = \varepsilon_{eff} \Delta x_r \quad , \tag{3.3}$$

where $\Delta x_r$ is the change on the RNA extension when the molecule unfolds and $\varepsilon_{eff}$ is the effective stiffness of the whole experimental system given by Eq. 2.5. The force difference (Eq. 3.3) is a consequence of the particular design of experimental setup (Fig.1). For longer handles (i.e. softer handles) or softer traps the value of the effective stiffness, and hence the force difference (Eq. 3.3), decreases. In this latter case the thermodynamic potential of the whole experimental system (Eq. 2.1) in ME approaches to the IFE case as shown in the Supplementary Materials (Fig. S1).

The overall effect of such fluctuations in the hopping kinetics is not straightforward because the F-U rates might be increased due to (i) and (iii), but also decreased due to (ii). In the limiting case of very soft handles, i.e. when the stiffness of the molecular construct $\varepsilon_x$ approaches zero (and therefore $\varepsilon_{eff}$ =0), all previous effects (i), (ii) and (iii) tend to disappear and the experimental conditions get closer to the IFE. However, the temporal and the spatial/force sensitivity are also expected to decrease for softer handles (1). The reason is two-fold. On the one hand, in order to measure the F-U rates, the response of the trapped bead must be faster than the F-U reaction, i.e. $\tau_b \ll k_{F-U}^{-1}$. The corner frequency of the trapped bead (given by the inverse of Eq. 2.6) becomes lower for softer handles, decreasing the temporal resolution of the experiment. On the other hand, in order to detect accurately enough the force/extension jumps that characterize the F-U transition the handles should be stiffer than the trap or $\varepsilon_x \geq \varepsilon_b$. Otherwise the signal-to-noise ratio (SNR) would become too low and the experimental signal given by the force/extension jumps could be masked by the handles (1). In the current experimental conditions $\varepsilon_b \in [0.035 - 0.1]$ pN/nm whereas $\varepsilon_x \in [0.15 - 1.5]$ pN/nm so the inequality $\varepsilon_x \geq \varepsilon_b$ is satisfied. It can be shown that when $\varepsilon_x \geq \varepsilon_b$ the magnitude of the force fluctuations described in (i) and (ii), is quite insensitive to the particular value of $\varepsilon_b$. The main effect of $\varepsilon_b$ is to modify the value of the force difference (iii) (Eq. 3.3), which is minimized by taking $\varepsilon_b$ as small as possible. Therefore, to get estimates closer to the intrinsic molecular rate softer traps should be used.

The resolution and limitations of the instrument are also important when acquiring the experimental data. In particular, measurements are sensitive to the bandwidth $B$ at which data are collected and to the time lag of the feedback mechanism, $T_{lag}$:

*(iv) Limited bandwidth:* If the bandwidth $B$ is not higher than the F-U rates, the time resolution of the measurement becomes too low to detect the F-U reaction and the measured kinetic rates will be affected.

*(v) Piezoelectric flexure stage:* In the CFM the force-feedback mechanism operates to compensate for the force difference given by Eq. 3.3. If $T_{lag} \ll k_{F-U}^{-1}$ then $\Delta \langle f_{RNA} \rangle = 0$ is verified and the feedback mechanism can efficiently keep the force constant. Otherwise, the feedback mechanism cannot maintain the force constant on timescales where the molecule folds/unfolds. In the latter case, the feedback mechanism leads to distorted rates. We call this the distortion effect (1).

If $B^{-1}$ and $T_{lag}$ are much shorter than $k_{F-U}^{-1}$, only the effects (i)-(iii) remain. By using longer (i.e. softer) handles and softer traps these effects are also minimized and the measured rates should approach the ideal molecular rates. For all the experimental setups we have investigated in this work the spatial/force resolution is high enough to detect the F-U reaction (1) and the condition $\tau_b \ll k_{F-U}^{-1}$ holds. Therefore, the optimal conditions to carry out measurements would be to use handles and traps as soft as possible within the limiting resolution imposed by the experimental setup (SNR>1, $\tau_b \ll k_{F-U}^{-1}$).

Even though current experiments (1) do not reveal the presence of intermediates of the F-U reaction, most of the kinetic effects observed in the experiments are not captured by a simple two-state model that does not include intermediate configurations. In fact, the two-state model only considers the dynamical effects (i) and (iii), which increase the RNA F-U kinetics as compared with the IFE case. To reproduce the observed dependence in the kinetics it is necessary to take into account the base-pair hopping effect (ii) in the dynamics. Therefore a multi-state model, as the one proposed here, is needed to capture the effect of the experimental setup on the measured kinetics.

## 4. Modeling hopping dynamics in the experimental modes.

In this section we study the RNA F-U kinetics under the experimental conditions by simulating the dynamics of the whole system in the PM and CFM. In Sec. 4.1 we describe the model we use for the F-U kinetics of the RNA hairpin. The simulation algorithm is presented in Sec. 4.2.

### 4.1 Kinetic model for folding/unfolding the RNA

To model the kinetics of the RNA hairpin we adapt the model by Cocco et al. (23) to our experimental setup; we assume the dynamics of the hairpin to be sequential (see Sec. 2.1). Therefore one-step transitions connect each configuration $n$ with its first nearest neighbors in the configurational space, $n+1$ and $n-1$. The dynamical process is then governed by the kinetic rates to go from $n$ to $n'$ with $n' = n-1, n+1$. This kinetic model is schematically depicted in Fig. 2. The evolution in time of the configuration $n$ of the hairpin is described by a set of coupled master equations:

$$\frac{dp_n(t)}{dt} = -(k_\rightarrow(n) + k_\leftarrow(n))p_n(t) + k_\leftarrow(n+1)p_{n+1}(t) + k_\rightarrow(n-1)p_{n-1}(t) \quad , \tag{4.1}$$

with $n = 0,...N$ and $k_\rightarrow(N) = 0$, $k_\leftarrow(0) = 0$, $p_{-1}(t) = 0$, $p_{N+1}(t) = 0$. The function $p_n(t)$ is the probability for the RNA molecule to be in the configuration $n$ at time $t$, $k_\rightarrow(n)$ is the transition rate to go from $n$ to $n+1$ at time $t$ and $k_\leftarrow(n+1)$ denotes the rate of the reverse reaction. The experimental system includes different elements such as the handles, the trapped bead and the RNA molecule, therefore the F-U kinetics is described by the rates associated to the transitions, $(x_b, x_{h_1}, x_{h_2}, x_r, n) \rightarrow (x_b', x_{h_1}', x_{h_2}', x_r', n')$ with $n' = n-1, n+1$ and $X_T - R_{b1} - R_{b2} = x_b + x_{h_1} + x_{h_2} + x_r = x_b' + x_{h_1}' + x_{h_2}' + x_r'$. Because the bead relaxes much slower than the handles and the ssRNA the kinetics of the hairpin is slaved to the relaxational dynamics of the bead. Consequently the kinetics rates can be factorized in two terms:

$$W[(x_b = z, n) \to (x_b = z', n')] = W_1(z \to z'/n) W_2(n \to n'/z'), \text{ with } n' = n+1, n-1, \quad (4.2)$$

where $W_1(z \to z'/n)$ is the transition rate to go from $x_b = z$ to $x_b = z'$ when the hairpin is in the conformation $n$ and $W_2(n \to n'/z')$ is the transition rate to go from $n$ to $n'$ when $x_b = z'$. The rates given in Eq. (4.2) must verify detailed balance:

$$\frac{W_1(z \to z'/n)}{W_1(z' \to z/n)} = \exp\left[\frac{-(z - \langle x_b(n)\rangle)^2 + (z' - \langle x_b(n)\rangle)^2}{2\langle \delta x_b^2(n)\rangle}\right],$$

$$\frac{W_2(n \to n'/z)}{W_2(n' \to n/z)} = \exp[\beta(G^0(n') + \Delta G_{x_b}(n') - G^0(n) - \Delta G_{x_b}(n))] \quad (4.3)$$

where $\beta = \frac{1}{k_B T}$, $\langle x_b(n)\rangle$ and $\langle \delta x_b^2(n)\rangle$ are the mean value and the fluctuations in the position of the trapped bead at the given value of the control parameter $X_T$ and at the given configuration of the hairpin $n$, $G^0(n)$ is the free energy of the configuration $n$ of the hairpin at zero force and $G_{x_b}(n)$ is the free energy contribution due to the handles and the ssRNA that are stretched a distance $x = X_T - x_b - R_{b1} - R_{b2}$. $G_{x_b}(n)$ is computed by using Eq. 2.4 as:

$$G_{x_b}(n) = \sum_{i=1,2} \int_0^{x_{h_i}(n)} f_{h_i}(x) dx + \int_0^{x_r(n)} f_r(x) dx \quad (4.4)$$

where $f_\alpha(x)$ with $\alpha = h_1, h_2, r$ corresponds to the equilibrium force extension relation as given by the WLC model (24,25). The variables $x_{h_1}(n)$, $x_{h_2}(n)$ and $x_r(n)$, that verify $x = X_T - x_b - R_{b1} - R_{b2} = x_{h_1} + x_{h_2} + x_r$, correspond to the extension of the handles 1 and 2 and the released ssRNA for the RNA configuration with $n$ opened bps. The choice of the opening and closing rates $k_\to(n) = W_2(n \to n+1/x_b)$ and $k_\leftarrow(n) = W_2(n \to n-1/x_b)$ is based on two assumptions (23): (i) The transition state corresponding to the formation-dissociation reaction of a given bp is located very close to the formed state. Therefore the opening rate $k_\to$ for a given bp depends on the particular bp and its neighbor (i.e., GC versus AU), but does not depend on the value of the control parameter $y$ (e.g. $X_T$ or $f$). (ii) The rate of closing $k_\leftarrow$ is independent of the sequence and is determined by the work required to form the bp starting from the dissociated state. The rates $k_\to(n), k_\leftarrow(n)$ are of the Arrhenius form and are given by:

$$k_\to(n) = k_a e^{-\beta \Delta G^0(n)}, \quad k_\leftarrow(n+1) = k_a e^{-\beta \Delta G_{x_b}(n)} ; \quad (4.5)$$

The constant $k_a$ is a microscopic rate that does not depend on the particular bp sequence and is equal to the attempt frequency of the molecular bond. The kinetic process defined by $k_\rightarrow$, $k_\leftarrow$ is of the activated type. The value of $k_\rightarrow$ is a function of the free energy difference $\Delta G^0(n) = G^0(n+1) - G^0(n)$ between the two adjacent configurations, $n$ and $n+1$. Whereas the value of $k_\leftarrow$ depends on the value of the control parameter $X_T$ and on the value of $x_b$, $x_{h_1}$, $x_{h_2}$ and $x_r$ through $\Delta G_{x_b}(n) = G_{x_b}(n+1) - G_{x_b}(n)$, where $G_{x_b}(n)$ is the free energy defined in Eq. 4.4. The choice of these rates has the advantage that there is only one free parameter, $k_a$, while the rest of parameters can be obtained from measured thermodynamics. This model is an extension of the one proposed by Cocco et al. (23), as given in Eq. 4.1, by considering the appropriate kinetic rates (Eq. 4.5) adapted to reproduce the experimental CFM and PM.

## 4.2 Monte Carlo simulation of hopping experiments

To simulate the hopping experiment we benefit from the large separation of timescales between the different elements of the system: $\tau_b \gg \tau_{handles}, \tau_{ssRNA}$ (Table 1). We consider that during the time of an iteration step in the simulation, $dt = 10^{-8}$ s, the handles and the ssRNA are in local equilibrium, but the bead in the trap is not. Note that the timescale of iteration is smaller than the relaxation time $\tau_{handles}$. However, we do not expect our results to change much by taking into account the microscopic dynamics of the handles, because the most important dynamical effect either in the simulations or the experiments comes from the bead in the trap. In fact, the bead is the element of the system with largest dissipation and slowest relaxation rate as compared to the elastic and bending modes of the handles and the ssRNA. In our simulation we implement the following algorithm:

- At each iteration step $dt$:

    1. The position of the bead trapped in the optical potential $x_b$ evolves according to the Langevin dynamics of an overdamped particle (22):

    $$\gamma \frac{dx_b}{dt} = -\varepsilon_b x_b + f_x(t) + \xi(t), \qquad (4.6)$$

    where $\varepsilon_b$ is the stiffness of the trap, $\gamma$ is the friction coefficient of the bead and $f_x$ is the force exerted by the molecular construct on the bead. $f_x$ is computed as the force needed to extend the molecular construct (the handles and the ssRNA) a distance $x = X_T - x_b - R_{b1} - R_{b2}$. The stochastic term $\xi(t)$ is a white noise with mean value $\langle\xi(t)\rangle = 0$ and variance $\langle\xi(t)\xi(t')\rangle = 2k_B T\gamma\delta(t-t')$. From the evolution of the bead position, given by Eq. 4.6, we obtain the instantaneous values of the molecular extension $x = X_T - x_b - R_{b1} - R_{b2}$ and force $f = \varepsilon_b x_b$.

    2. For a given extension $x = X_T - x_b - R_{b1} - R_{b2}$, we compute the equilibrium value of the extension of the handles and the released ssRNA for the configurations with $n$ and $n-1$ opened bps, by using the WLC model (24,25). We then compute the function $\Delta G_{x_b}(n)$ as given by Eq. 4.4.

3. We change the configuration of the hairpin from $n$ to $n$, $n+1$ or $n-1$ with probabilities [1- $k_{\rightarrow}(n)dt$ - $k_{\leftarrow}(n)dt$], $k_{\rightarrow}(n)dt$ and $k_{\leftarrow}(n)dt$ respectively, where the rates $k_{\rightarrow}$ and $k_{\leftarrow}$ are defined in Eq. 4.5.

4. Return to 1.

- We average the instantaneous data over a bandwidth $B$.

- In the CFM, at every 1 ms of time, we increase (decrease) the value of the total end-to-end distance $X_T$ by 0.25 nm if the measured force differs by more than 0.1 pN below (above) the set point force value at which the feedback mechanism operates.

## 5. Reaction rates from the hopping traces.

In this section we compute the rates of the F-U reaction from the hopping traces corresponding to the CFM and PM simulations. We then compare them with the experimental results (1). We use the free energy parameters given in (33,34,35) to compute the free energy landscape at zero force $G^0(n)$ of the P5ab hairpin at 25 °C and in 1M NaCl *. We consider that the mechanical response of the handles and the ssRNA is characterized by a persistence length ($P$) and contour length ($L$) equal to $P_h$ = 10 nm, $L_h$ = 0.26 nm/bp for the handles and $P_{ssRNA}$ = 1 nm, $L_{ssRNA}$ = 0.59 nm/base for the ssRNA. In order to analyze the effect of the instrument on the measured rates we study different experimental setups by considering handles of several lengths and optical traps characterized by different stiffness.

Figs. 3 and 4 show examples of CFM and PM traces obtained from the simulations of the experimental system. The distributions of lifetimes obtained either from the experimental traces or from the simulations have an exponential decay (Supplementary Materials, Fig. S2), as expected for a two-state system. To extract the rates of the F-U reaction in each mode we have analyzed the simulated data using the same methods as for the experimental data (1). We then compare these rates with the experimentally measured rates presented in our companion paper (1). The value of the free parameter $k_a$ is chosen to optimize the fit between the rates extracted from the simulation traces and the ones measured in the experiments. For this fit we used the PM data as explained in the Supplementary Materials (Fig. S3). Notice that the value of $k_a$ fixes the timescale unit of the simulation allowing us to establish the connection between the real microscopic dynamics of the molecule and the mesoscopic description. We get the characteristic bp attempt frequency, $k_a = 2.3 \cdot 10^6$ Hz *. By solving the master equation (Eq. 4.1) for the F-U reaction in the IFE (23), we also get an estimate for the intrinsic molecular rate $k^0_{F-U}$ at the critical force. We obtain $k^0_{F-U}$ = 13 Hz. In what follows we compare this value with the measured rates in the CFM and PM in order to infer the optimal conditions to obtain rates as close as possible to the intrinsic molecular rate $k^0_{\text{F-U}}$.

## 5.1 Constant-force mode (CFM)

In Fig. 5 we show the values of the CFM critical rates, $k^c_{\text{CFM}}$, obtained from the simulations as a function of the length of the handles (filled symbols) compared with the experimental ones (1) (empty symbols connected by lines). The agreement between the experimental and simulation results in the CFM is reasonable. The analysis done in Sec. 3.2 predicts that the measured critical rates should converge to the value of the intrinsic molecular rate $k^0_{\text{F-U}}$ for softer handles, i.e. longer handles. This is true when the instrument has enough time resolution to resolve the force/extension jumps, i.e. $k_{\text{F-U}}^{-1} \gg T_{\text{lag}}, B^{-1}$. However, we do not observe this convergence, neither in the simulations nor the experiments (Fig. 5), probably because in our instrument $k_{\text{F-U}}^{-1} \sim T_{\text{lag}} \sim 0.1$ s, such that the condition $k_{\text{F-U}}^{-1} \gg T_{\text{lag}}$ is not satisfied. In this situation, the measured rates highly depend on the bandwidth and on the criteria used to analyze the data, i.e. the so-called distortion effect discussed in the companion paper (1). We think that the non-convergence of the measured rates for long handles to the intrinsic molecular rate $k^0_{\text{F-U}}$ arises from distortion effects due to the finite response time of the instrument, $T_{\text{lag}}$. To validate this hypothesis and to obtain better estimates for the rates, we propose to use the PM data to extract the PM critical rate $k^c_{\text{PM}}$ (see Sec. 3.1). In the PM there is no feedback mechanism; therefore PM data does not suffer from the distortion effect. Also, by using a bandwidth high enough, i.e. $B \gg k_{\text{F-U}}$, we waive the dependence of the measured rates on the bandwidth. Therefore, the PM critical rate $k^c_{\text{PM}}$ should provide a better estimate of the F-U rate at the critical force.

## 5.2 Passive mode (PM)

From the PM data, we extract the PM rates. By doing numerical simulations at different values of $X_T$, we obtain the PM folding and unfolding rates at different forces. As shown in Fig. 6 the logarithm of the PM folding and unfolding rates as a function of the force fits well to a straight line, as predicted by the Kramers-Bell theory for two-state systems (18). The experimental measured rates show the same dependence on the force as the simulation results (Fig. 6), suggesting that the model proposed predicts well the location of the transition state (17). The PM critical rate $k^c_{\text{PM}}$ is obtained from the intersection of the linear fits to the computed data for $\ln(k_U)$ and $\ln(k_F)$ as a function of the force. In Fig. 7 we show the measured PM critical rates from the simulation traces (filled symbols) as well as the experimental results (1) (empty symbols connected by lines) as a function of the length of the handles. Two sets of data at the trap stiffness $\varepsilon_b = 0.1$ pN/nm and $\varepsilon_b = 0.035$ pN/nm are shown. Both experimental and simulation results agree pretty well. The bandwidth used, $B = 1$ KHz, is much greater than $k_{\text{F-U}}$. Hence, the time resolution is sufficient to follow the F-U reaction, and the measured rates are not affected by the average of the data over the time window $B^{-1}$. Better estimates are obtained for the softer trap case as expected.

Finally, in Fig. 8 we compare the critical PM (filled symbols connected by lines) and CFM rates (empty symbols connected by lines) measured in the experiments. The discrepancy between the critical rates $k_{PM}^c$ and $k_{CFM}^c$ is larger for the stiffest trap results ($\varepsilon_b$ = 0.1 pN/nm, upper panel) and the longest handles, case in which distortion effects in the CFM are more important (1). Moreover, the values of the rates $k_{PM}^c$ obtained from the PM data in both experimental setups (upper and lower panels) increase for longer handles and show a tendency to approach to the ideal molecular rate value of 13Hz as expected (see Sec. 3.2). These results confirm our initial expectations that, then $k_{F-U}^{-1}$ is of the order of $T_{lag}$, the measured rates in the CFM are strongly affected by the distortion effect.

### 5.3 The quality factor Q

To compare different estimates for the critical rates, it is useful to define a parameter that characterizes the reliability of the measurement. We define the quality factor $Q$ as the relative difference between the measured rate ($k_{est.}$) and the intrinsic molecular rate $k_{F-U}^0$:

$$Q = 1 - \left| \frac{k_{est.} - k_{F-U}^0}{Max(k_{est.}, k_{F-U}^0)} \right| \begin{cases} Q = \frac{k_{est.}}{k_{F-U}^0} & \text{if } k_{F-U}^0 > k_{est.} \\ Q = \frac{k_{F-U}^0}{k_{est.}} & \text{if } k_{F-U}^0 < k_{est.} \end{cases} \quad (5.1)$$

As a compendium of all the results, we show in Fig. 9 the value of $Q$ obtained for the different estimates for the critical rates as extracted from the experimental data. The factor $Q$ is shown as a three-dimensional plot as a function of the length of the handles and the trap stiffness. We show 3 surfaces, each corresponding to a different estimate of the rates: the CFM critical rates ($k_{CFM}^c$) for two different values of the bandwidth, and the PM critical rates ($k_{PM}^c$).

Depending on the RNA molecule (sequence, length, folding and unfolding rates) and the characteristics of the experimental setup (trapped bead, handles, feedback time lag and bandwidth), the quality factor of each estimate may change. As a general result we infer that better measurements are obtained for softer traps and longer handles as long as the transition signal is detectable. For fast hoppers (which have F-U rates which are not much slower than the force-feedback frequency, as happens in our study of the P5ab hairpin where $k_{F-U}^{-1} \sim T_{lag}$, ~0.1s), PM rates provide better estimates than CFM rates. On the other hand, for slow hoppers, the PM becomes impractical due to the presence of drift effects. In the latter case, the CFM is efficient and the CFM critical rates should be a good estimate for the intrinsic molecular rates.

### 6. Summary and Conclusions

In this work we have introduced a mesoscopic model for the study of the folding/unfolding (F-U) force-kinetics of RNA hairpins in hopping experiments using optical tweezers. The model incorporates the different elements of the experimental setup (bead, handles and RNA sequence) and limitations of the instrument (time lag of the constant-force mode and finite bandwidth). We carry out numerical simulations of the proposed model and compare them with hopping experiments in the P5ab RNA hairpin reported in our companion article (1). This analysis allows us to extract the value of the microscopic attempt frequency $k_a$ for the dissociation kinetics of individual base-pairs. The estimate for $k_a$ is then used to extract the intrinsic molecular rate for the RNA hairpin, $k_{F-U}^0$. We then compare the estimate of the intrinsic molecular rate with the values for the different rates (constant-force mode (CFM) and passive mode (PM) rates) obtained under different experimental conditions. The goal of the research is to infer the optimal conditions to extract the intrinsic molecular rate of the RNA molecule using data obtained in the different experimental modes: passive and constant-force. We have considered different values of the stiffness of the trap and different lengths of the handles. Due to the complexity of the system the quality factor $Q$ (defined as the relative difference between the measured rate and the intrinsic molecular rate) will critically depend on various parameters of the instrument (experimental setup and the instrumental limitations) and the molecule.

Through our analysis we are able to find the optimum experimental conditions to measure hopping rates. Even though our study has been carried out for an RNA hairpin with a fixed sequence in an optical tweezers setup, the methodology and rationale presented here can be applied to other experimental setups, such as dumbbell dual-trap optical tweezers (21,32,36,37), other acid nucleic sequences, or proteins (12). Our main conclusions can be summarized as follows:

- *Trap.* For all experimental modes it is advisable to use traps as soft as possible [$\varepsilon_b \leq 0.1$ pN/nm]. In particular, in order to detect the force/extension jumps that characterize the F-U transition the trap should not be stiffer than the handles.
- *Handles.* For all experimental modes it is advisable to use handles as long as possible within the resolution limit of the instrument [$3$ Kbp $\leq L_h \leq 10$ Kbp]: (i) the SNR of the extension/force signal must be large enough to follow the F-U reaction and (ii) the corner frequency of the bead (equal to the inverse of its relaxation time) must be much higher than the F-U rate of the hairpin (see the discussion in (1)).
- *Bandwidth.* For all experimental modes it is advisable that the bandwidth of data collection is as large as possible.
- *Force-feedback frequency.* In the CFM it is important that the frequency of the force-feedback mechanism is as high as possible. In particular, the force-feedback frequency must be higher than the F-U rate, otherwise distortion effects are big and the force-feedback mechanism becomes inefficient. If the latter restriction is not satisfied (as happens in our study of the P5ab hairpin where $k_{F-U}^{-1} \sim T_{lag} \sim 0.1$ s) then PM rates provide better estimates than CFM rates. For our experimental setup the CFM should be more efficient in studying slow RNA hoppers (e.g. RNA molecules with tertiary interactions) that satisfy $k_{F-U}^{-1} \gg T_{lag}$.

To measure rate constants up to 100 Hz with certain accuracy for RNA molecules of about 20 bps long, requires traps softer than 0.1 pN/nm, handles longer than 2 Kbp but shorter than 15 Kbp, and bandwidth and force-feedback frequency of 1 KHz or higher. In all cases we studied in this work the different estimates for the rates are of the same order of magnitude as the intrinsic molecular rate. Optical tweezers are, thus, a very useful single-molecule technique to infer the values of the force dependent F-U kinetic rates of biomolecules. Future design of optical tweezers with higher spatial resolution and higher frequency force-feedback mechanisms will provide better instruments to characterize the F-U kinetics of biomolecules.

## TABLES

**Table 1:** Different characteristic timescales of the system shown in Fig.1: relaxation time of the bead in the trap $\tau_b$; the relaxation time associated to the longitudinal modes of the handles and the ssRNA denoted by $\tau_{handles}$ and $\tau_{SSRNA}$ respectively; the RNA hairpin folding-unfolding time $k_{F-U}^{-1}$, bp breathing time $k_{bp}^{-1}$; and intrinsic times of the instrument: average sampling time $B^{-1}$ and time lag of force feedback mechanism $T_{lag}$.

| $\tau_b$ [ms] | $\tau_{handles}$ [ms] | $\tau_{SSRNA}$ [ms] | $k_{F-U}^{-1}$ [ms] | $k_{bp}^{-1}$ [ms] | $B^{-1}$ [ms] | $T_{lag}$ [ms] |
|---|---|---|---|---|---|---|
| $1-10^{-2}$ | $10^{-3}-10^{-5}$ | $3.5 \cdot 10^{-6}$ | $10^4-10$ | $10^{-3}-10^{-5}$ | $10^{-2}-1$ | $10^2$ |

## FIGURE CAPTIONS

**Figure 1:** Schematic picture of the model for the experimental setup used in the manipulation of RNA molecules. We show the configurational variables of the system $x_b$, $x_r$, $x_{h1}$ and $x_{h2}$ which are the extensions of each element (trapped bead, RNA molecule and handles respectively) along the reaction coordinate axis (i.e. the axis along which the force is applied). $X_T$ is the end-to-end distance of the whole system, i.e. the distance between the center of the optical trap and the tip of the micropipette. The optical potential is well described by a harmonic potential of a one-dimensional spring of stiffness $\varepsilon_b$ and equilibrium position at $x_b = 0$.

**Figure 2:** Schematic representation of the kinetic model for the RNA hairpin. The model assumes that the dynamics of the folding and unfolding of the hairpin is sequential. Therefore each intermediate configuration $n$ is only connected to its first neighbors $n+1$ and $n-1$, where $n$ represents the number of sequential bps unpaired from the opening of the helix. The kinetic rates to go from $n$ to $n-1$ or $n+1$ govern the F-U dynamical process.

**Figure 3:** Extension traces for P5ab hairpin with 3.2 Kbp handles in CFM from the simulations for two different trap stiffness $\varepsilon_b \approx 0.1$ pN/nm (upper panel) and $\varepsilon_b \approx 0.035$ pN/nm (lower panel). The bandwidth used is 200 Hz.

**Figure 4:** Force traces for P5ab hairpin with 3.2 Kbp handles in PM from the simulations for two different trap stiffness $\varepsilon_b \approx 0.1$ pN/nm (upper panel) and $\varepsilon_b \approx 0.035$ pN/nm (lower panel). The bandwidth used is 1 KHz. We show the mean forces, $f^F$ and $f^U$, in the upper and lower parts of the square-like-sign force traces, corresponding to the forces at the folded and unfolded states, respectively. Note that the value of such forces, $f^F$ and $f^U$, is higher than the ones measured in experiments (1) by about 1-1.5 pN. This discrepancy is consistent with the fact that the free energy parameters for P5ab used in simulations (32,33,34) corresponds to higher salt concentrations than the experimental ones*.

**Figure 5:** CFM critical rates as a function of the length of the handles from the experiments (empty symbols connected by lines) and simulations (filled symbols) for two different values of the trap stiffness $\varepsilon_b$ = 0.1 pN/nm (upper panel) and 0.035 pN/nm (lower panel). Results obtained by using different bandwidth $B$ = 10, 50 and 200 Hz (circles, squares and triangles, respectively) are shown. The molecular rate $k_{\text{F-U}}^0$ (dotted line) is also shown in the bottom panel for comparison. Better results are obtained for the softest trap $\varepsilon_b$ = 0.035 pN/nm where distortion effects are less important.

**Figure 6:** The logarithm of the PM folding (in blue) and unfolding (in red) rates as a function of force from experiments (empty circles) and simulations (filled circles). The folding and unfolding lines from simulations have been shifted by 1.5 pN*. Straight lines are the linear fits to the data from the simulation. The intersection point between the folding and unfolding lines gives the value of the PM critical rate.

**Figure 7:** PM critical rates as a function of the length of the handles measured in PM from experiments (empty circles connected by lines) and simulations (filled symbols) for two different values of the trap stiffness $\varepsilon_b$ = 0.1 pN/nm (squares) and 0.035 pN/nm (circles). The bandwidth (1 KHz) is much larger than the characteristic frequency of the F-U reaction. The intrinsic molecular rate $k_{\text{F-U}}^0$ (dotted line) is shown for reference. The PM rates show a tendency to approach to the value of $k_{\text{F-U}}^0$ for long handles as the analysis done in Sec. 3.2 predicts. Better results are also obtained for the softest trap $\varepsilon_b$ = 0.035 pN/nm. The agreement between the experiments, simulations and theory is good.

**Figure 8:** We compare the experimental CFM critical rates (empty symbols connected by lines) with the experimental PM critical rates (filled diamonds connected by lines) for two different trap stiffness $\varepsilon_b$ = 0.1 pN/nm (upper panel) and $\varepsilon_b$=0.035pN/nm (lower panel). CFM results with bandwidths of 10 Hz, 50 Hz and 200 Hz are shown in triangles, squares and circles respectively.

**Fig 9:** We show the quality factor $Q$ (defined as the closeness between the measured rate and the intrinsic molecular rate) obtained from different measured critical rates in experiments as a function of the length of the handles and the trap stiffness. For the CFM experimental results we show the $Q$ corresponding to the CFM critical rates at two different values of the bandwidth, $B$=200Hz (red) and $B$=10Hz (black). In blue it is also shown the $Q$ for the PM critical rates extracted from PM experiments. Generally, better measurements (higher Q values) are obtained from softer traps and longer handles.

# FIGURES

**FIG. 1:**

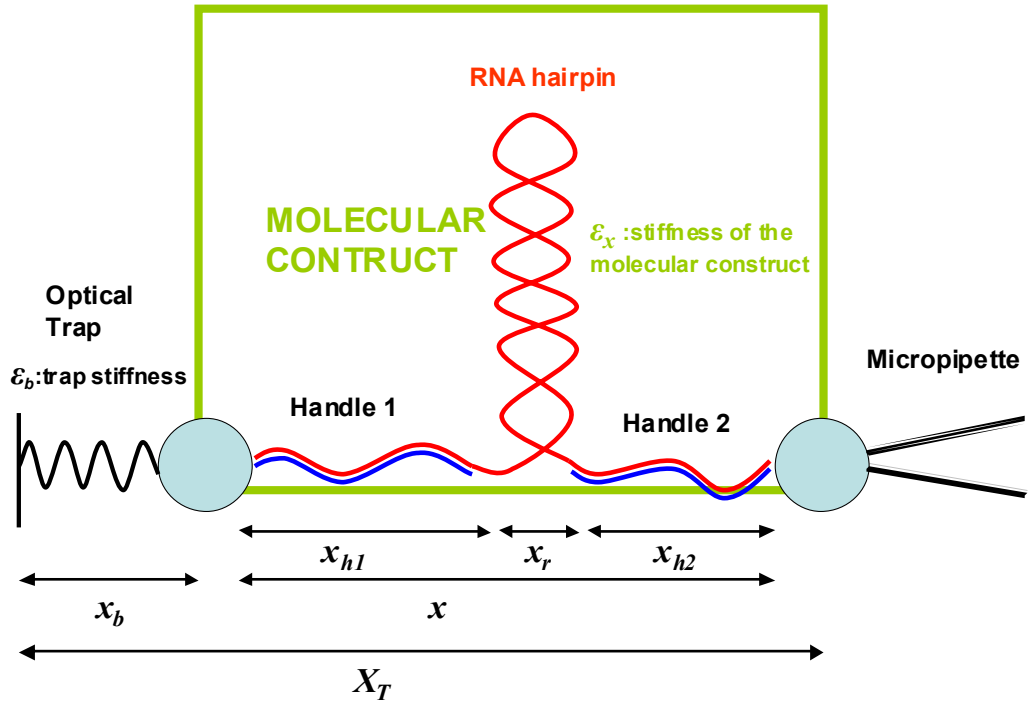

**FIG. 2:**

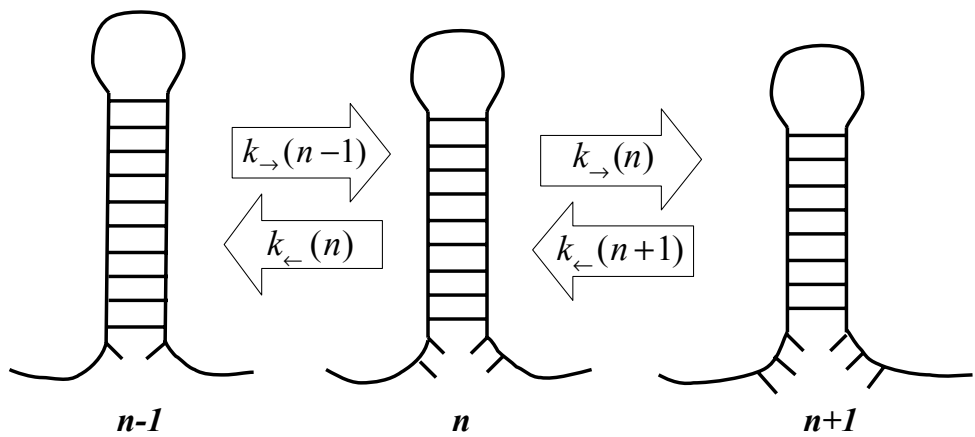

**FIG. 3:**

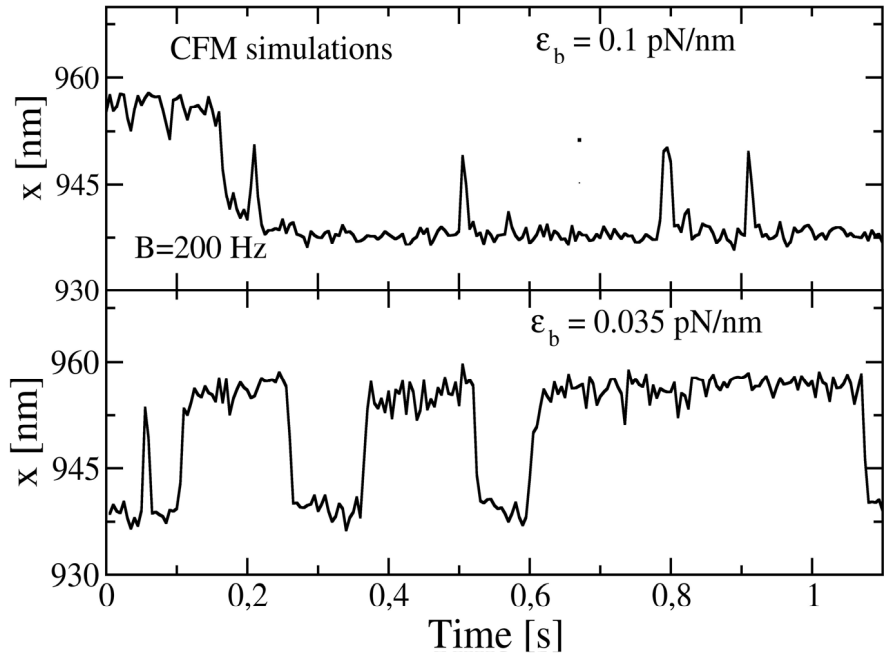

**FIG. 4:**

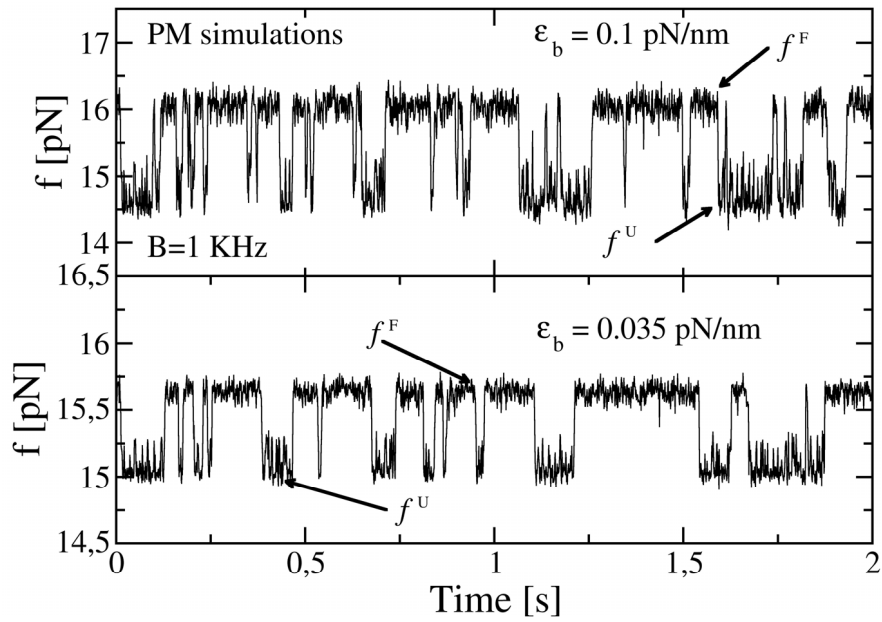

**FIG. 5:**

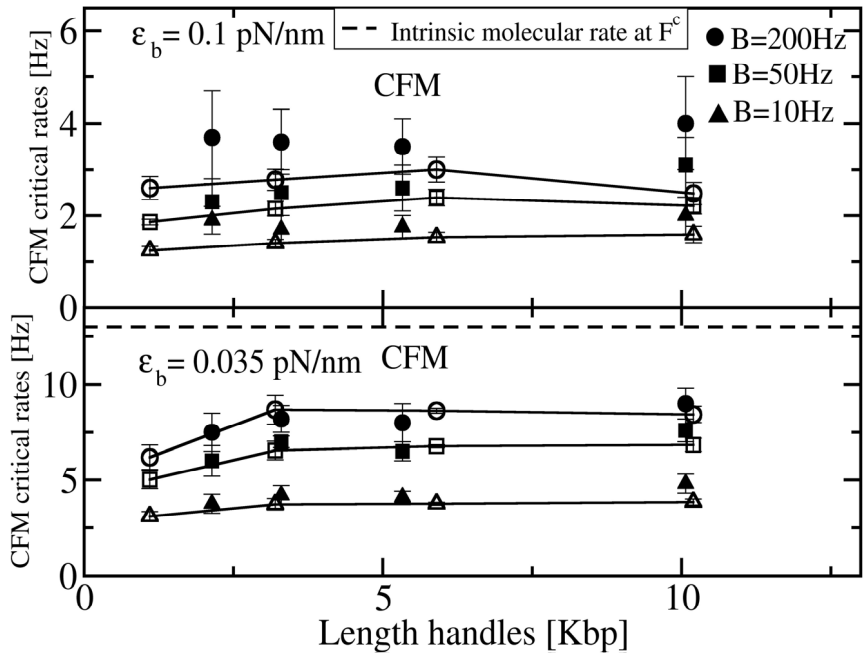

**FIG. 6:**

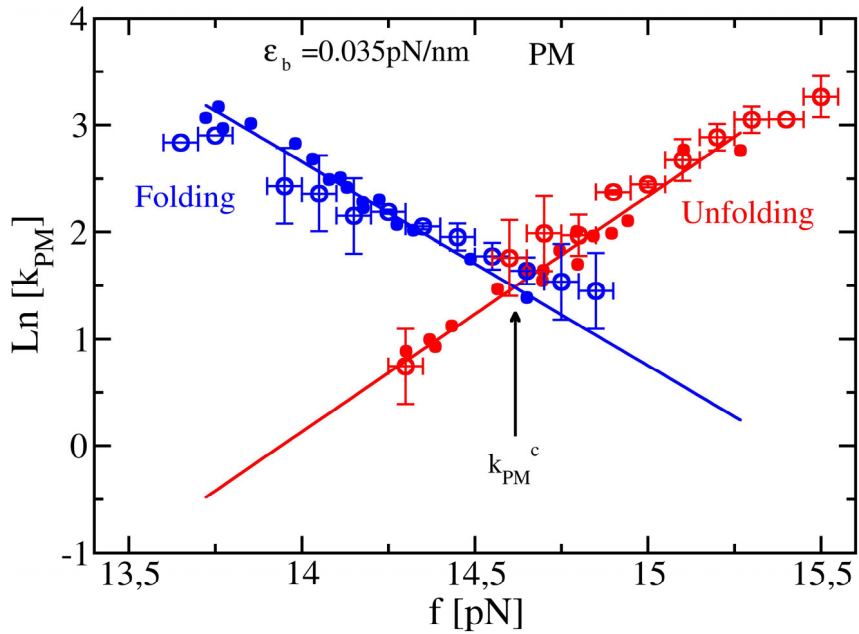

**FIG. 7:**

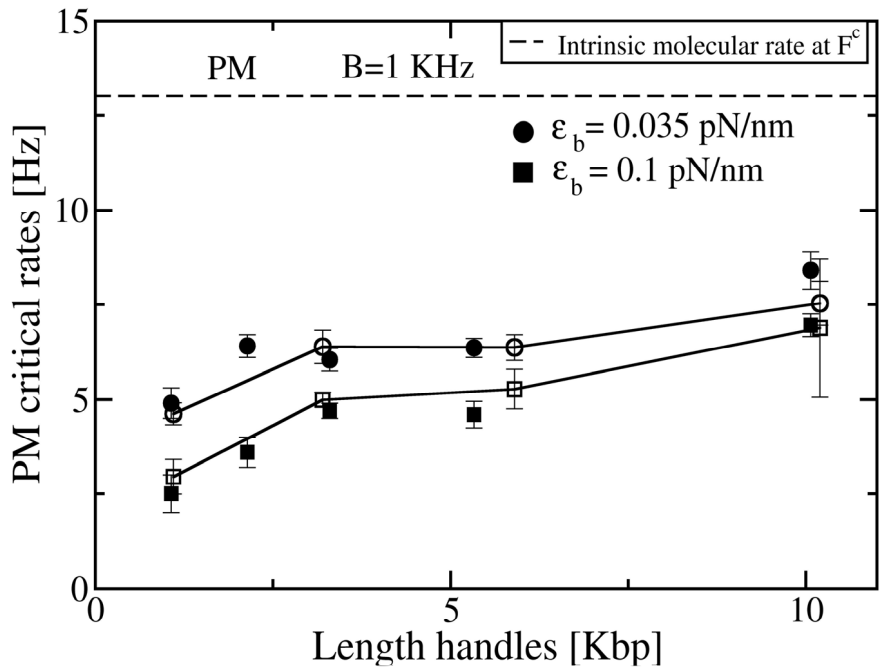

**FIG. 8:**

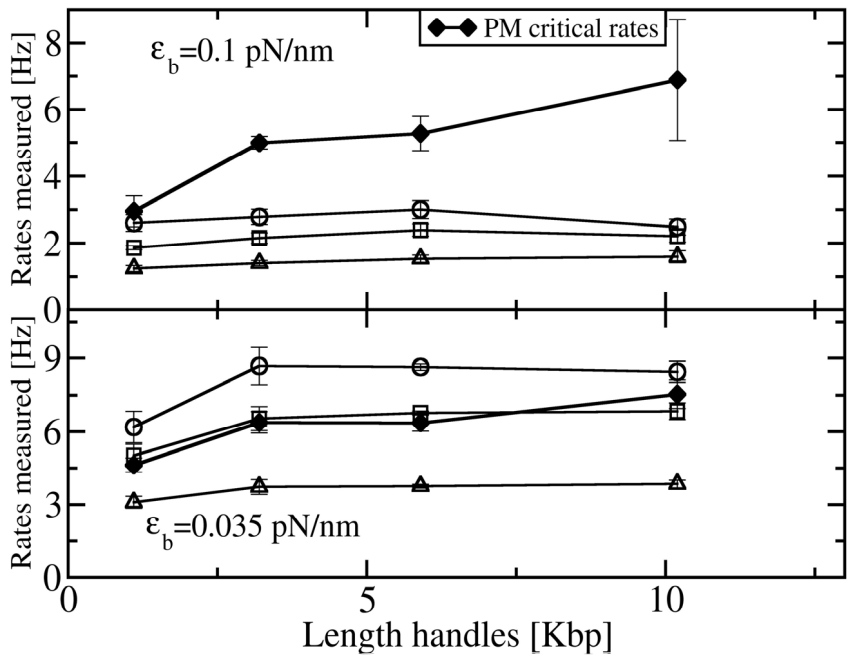

**FIG. 9:**

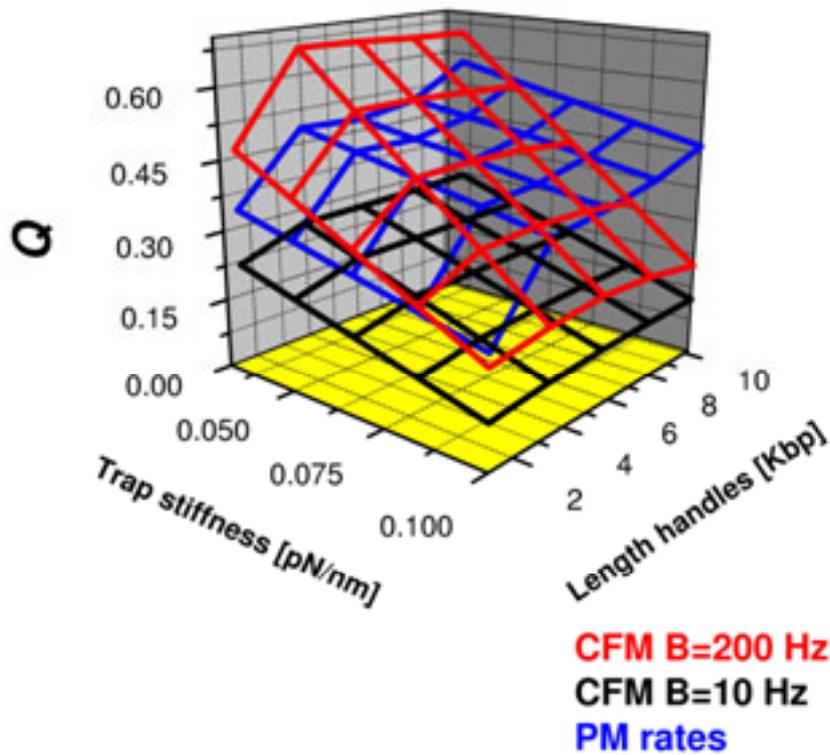

---

[*] Our experiments were performed at 250mM NaCl while our simulations used the free energy parameters obtained from (33,34,35) at 1M NaCl. The presence of salt in the solvent stabilizes folded conformations of RNA molecules due to the larger screening of the electrostatic repulsion between the phosphates groups. Therefore, the RNA native structure at higher salt concentrations has a lower free energy (i.e it is more stable) and the critical value of the force for the folding/unfolding reaction is larger. This is in agreement with the fact that the values of the critical force that we obtain in our simulations are about 1-1.5 pN above the ones measured in the experiments. In order to reproduce the thermodynamic properties of the RNA molecule from our simulations we must shift the forces by 1-1.5 pN downward. In addition, the salt concentration might also affect the F-U kinetics. The value we estimate for the attempt frequency $k_a$ by fitting simulations and experiments already incorporates the salt correction.

# Supplementary Materials

Thermodynamics of the system

In Fig. S1 we show the thermodynamic potential given by Eq. 2.1 for different values of the stiffness of the trap and the length of the handles. Calculations have been done in the P5ab RNA molecule at the critical value of the extension $X_T^c$ where the F and U states are equally populated. The different contributions to the total free energy are $G^0(n)$ (estimated using the free energy parameters given in (33,34,35)), the elastic contributions $W_\alpha(z)$ (with $\alpha = h_1, h_2, r$), given by the work necessary to stretch the handles and the ssRNA (Eq. 2.4, using the WLC model with parameters given in Sec. 5), and the potential energy of the bead in the trap (described by a harmonic potential of stiffness $\varepsilon_b$). The figure shows how the kinetic barrier separating the F and U states increases when the value of the effective stiffness $\varepsilon_{\text{eff}}$ decreases, i.e. softer traps and longer handles, approaching to the IFE case.

**Figure S1:** The thermodynamic potential $G(X_T^c, n)$ defined in Eq. 2.1 for the P5ab RNA molecule at the critical value of the extension $X_T^c$ for different characteristics of the system: 1 Kbp handles with different trap stiffness (0.1 pN/nm, red dashed line; 0.035 pN/nm, green dashed line), and 10 Kbp handles with trap stiffness 0.1 pN/nm (dashed brown). The continuous blue line corresponds to the thermodynamic potential $G(F^c, n)$ in the ideal force ensemble (IFE). $G(X_T^c, n)$ approaches $G(F^c, n)$ for softer traps or longer handles, i.e. when the value of the effective stiffness $\varepsilon_{\text{eff}}$ approaches zero.

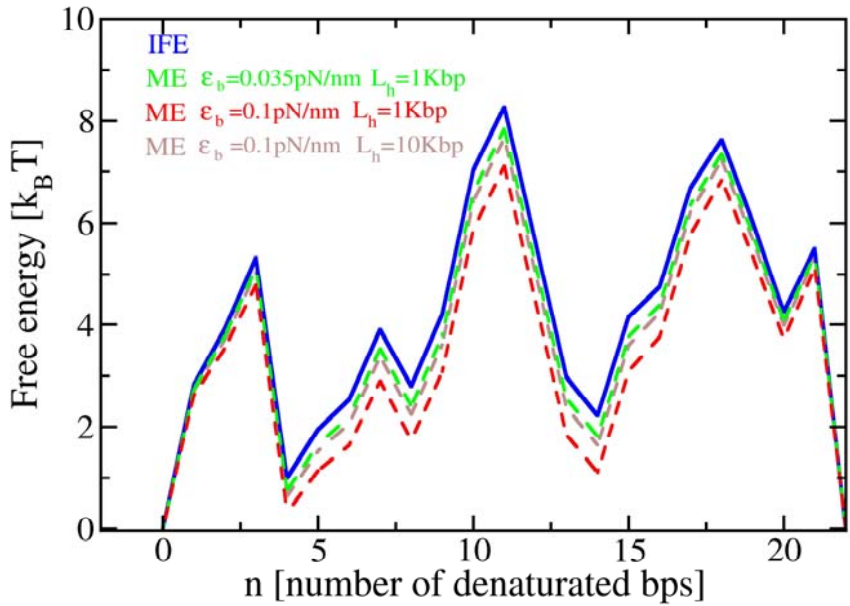

## Experimental and simulated distribution of lifetimes

From the simulations, we extract the distribution of lifetimes for the F and U state, $\tau_F$ and $\tau_U$. The distributions obtained from both the experiments and simulations fit well to an exponential (as shown in Fig. S2). This is a characteristic signature of two-state systems. The rates of the reaction are equal to the inverse of the mean lifetimes obtained either by fitting these distributions to an exponential or by taking the average of the measured lifetimes ($\tau_F$ or $\tau_U$).

**Figure S2:** Distribution of unfolding times obtained from the PM experimental traces (left panel) as compared with the PM simulations results (right panel) for trap stiffness 0.035pN/nm and for handles of 3.2 Kbp long. The bandwidth used is 1 KHz.

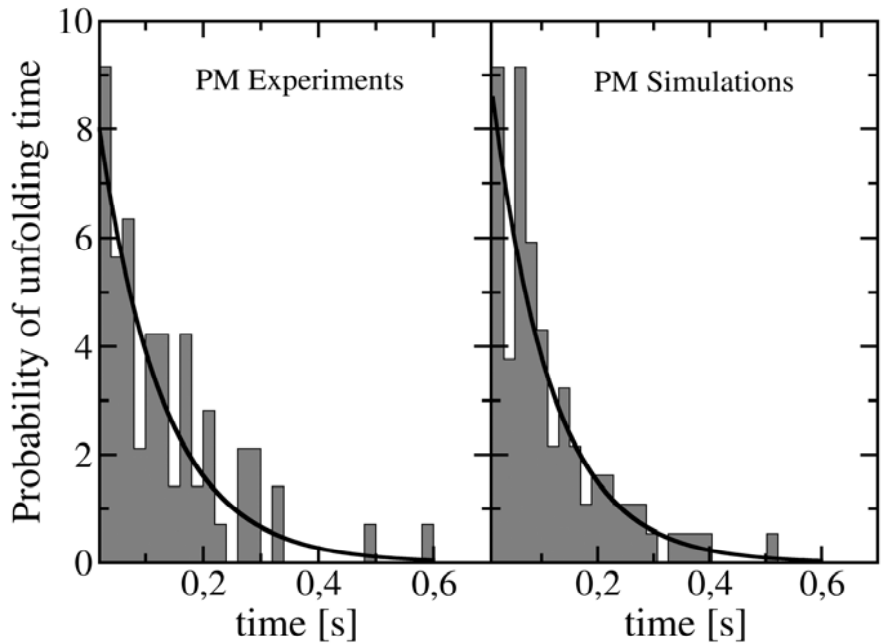

## Apparent rates

The apparent rates are the F-U rates (Eq. 2.8) measured in the PM when the distance $X_T$ is held fixed. We will consider the critical rate $k^c_{app}$ which is the F-U rate measured at the critical value $X_T^c$, where both states, F and U, are equally populated. When the hairpin unfolds the force drops whereas when the hairpin folds the force rises. In both cases the probability of folding and unfolding increases as compared with the CFM. Therefore we expect the apparent critical rates to be larger than the CFM critical rates, $k^c_{app} > k^c_{CFM}$. This is consistent with what has been experimentally measured (Figs. S3 and 5).

The value of the free parameter $k_a$ is estimated by fitting the rates extracted from the simulations to the experimental ones. For the fit we use the apparent rates because these are the ones less affected by the instrument. In Fig. S3 we show the measured rates from the simulation traces (filled symbols) as well as the experimental results (1) (empty symbols connected by lines) as a function of the length of the handles. Two sets of data at the trap stiffness, $\varepsilon_b = 0.1$ pN/nm and $\varepsilon_b = 0.035$ pN/nm are shown. Both experimental and simulation results agree pretty well. The bandwidth used, $B = 1$ KHz, is much greater than $k_{F-U}$. Hence, the time resolution is sufficient to follow the F-U reaction, and the measured rates are not affected by the average of the data over the time window $B^{-1}$. Consistently with the analysis done in Sec. 3.2, the results presented in Fig. S3 show that the measured critical rates approach the intrinsic molecular rate $k^0_{F-U}$ as handles extend. Moreover, better estimations are obtained for the softer trap case as expected. Note that, in the experimental setup studied here the values of the critical apparent F-U rates measured in the PM are closer to the value of the intrinsic molecular rates than the values of either the CFM critical rates or the PM critical rates (Fig. 8). The fact that the apparent critical rates are the best estimates for the intrinsic molecular rates is just a consequence of the compensation between the different dynamical effects in the PM. However, this may not be a general result.

**Figure S3:** Apparent critical rates as a function of the length of the handles measured in PM from experiments (empty symbols connected by lines) and simulations (filled symbols) for two different values of the trap stiffness $\varepsilon_b = 0.1$ pN/nm (squares) and 0.035 pN/nm (circles). The bandwidth (1 KHz) is much larger than the characteristic frequency of the F-U reaction. The intrinsic molecular rate $k^0_{F-U}$ (black dotted line) is shown for reference. The apparent rates approach to the value of $k^0_{F-U}$ for long handles as the analysis done in Sec. 3.2 predicts. Better results are also obtained for the softest trap $\varepsilon_b = 0.035$ pN/nm. The agreement between the experiments, simulations and theory is good.

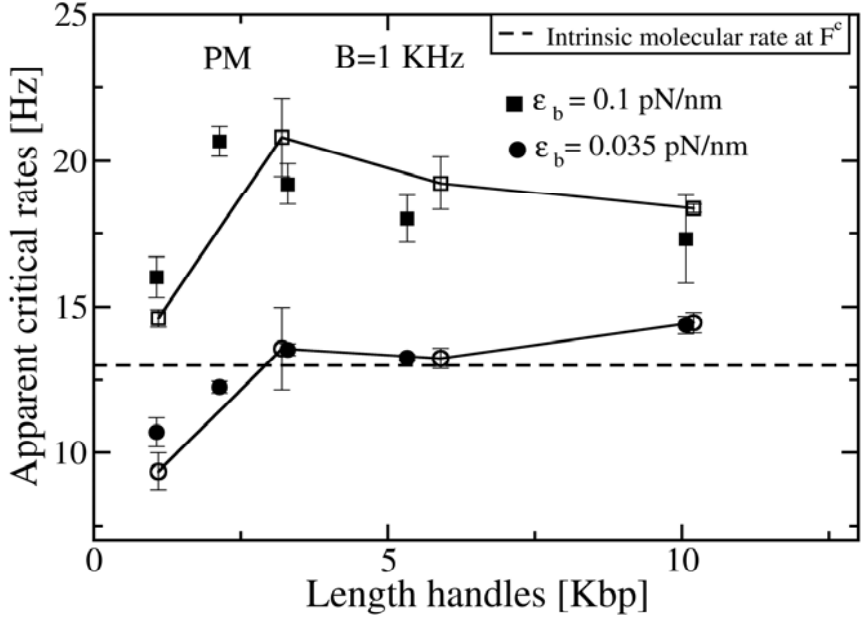

### Effect of the fluctuations described in (i) in the F-U kinetics of a hairpin

The force exerted upon the RNA is not controlled but fluctuates due to the bead force fluctuations (Eq. 3.1). In the frame of the two-state model we can write the equations for the probability densities, $p_F$ and $p_U$, for the molecule to be in the F and U state at a certain value of the force $f$ and at time $t$ respectively as:

$$\frac{dp_F(f,t)}{dt} = \int df' K^t_{\leftarrow}(f|f') p_U(f',t) - \int df' K^t_{\rightarrow}(f|f') p_F(f',t),$$

$$\frac{dp_U(f,t)}{dt} = \int df' K^t_{\rightarrow}(f|f') p_F(f',t) - \int df' K^t_{\leftarrow}(f|f') p_U(f',t), \quad (A1)$$

with the normalitzation condition:

$$P_U(t) + P_F(t) = 1, \quad \text{where} \quad P_\sigma(t) = \int df p_\sigma(f,t), \quad \text{with } \sigma = F,U. \quad (A2)$$

The functions $K^t_{\rightarrow}$ and $K^t_{\leftarrow}$ are the kinetics rates to jump from F to U and from U to F states, respectively, going from $f'$ to $f$ at $t$. The master equation Eq. A1 for the probabilities $p_F$ and $p_U$ is solved by discretizing in time and taking a discretization time $\Delta t$ smaller than the relaxation time for the F-U process, $k_{F-U}^{-1}$. For the system we are studying $k_{F-U}^{-1}$ and the relaxation time $\tau_b$ for the bead in the trap are of the order of 0.1 s and $10^{-5}$ - $10^{-3}$ s respectively (Table 1) . Then we can use a discretization of time large enough to allow for the relaxation of the bead. By considering $\Delta t > \tau_b$, the forces at initial and final time, $t$ and $t + \Delta t$, are decorrelated and we can rewrite the rates $K^t_\rightarrow$ and $K^t_\leftarrow$ as functions only of the final state at force $f$:

$$K^t_\rightarrow(f|f') = \rho(f,t)k_\rightarrow(f), \quad K^t_\leftarrow(f|f') = \rho(f,t)k_\leftarrow(f), \tag{A3}$$

where $\rho(f,t)$ is a normalized Gaussian distribution of mean $\langle f \rangle$ and variance $\delta f^2 = \langle \delta f^2_{RNA} \rangle = \delta^2$ given by Eq. 3.1. The functions $k_\rightarrow$ and $k_\leftarrow$ correspond to the rates in the IFE, i.e. $k_\rightarrow(f) = k^0 \exp(\beta f x_1)$ and $k_\leftarrow(f) = k^0 \exp(\beta(\Delta G - f x_2))$, where $x_1$ and $x_2$ are the distances form the F and U state to the transition state along the reaction coordinate axis and $\beta = \dfrac{1}{k_B T}$. By integrating out the final force $f$ in Eq. A1 we find:

$$\frac{dp_F(t)}{dt} = k'_\leftarrow(\langle f \rangle, \delta) p_U(t) - k'_\rightarrow(\langle f \rangle, \delta) p_F(t)$$
$$\frac{dp_U(t)}{dt} = k'_\rightarrow(\langle f \rangle, \delta) p_F(t) - k'_\leftarrow(\langle f \rangle, \delta) p_U(t), \tag{A4}$$

where the new effective rates $k'$ are given by:

$$k'_\rightarrow(\langle f \rangle, \delta) = k_\rightarrow(\langle f \rangle)\exp(\beta^2 x_1^2 \delta^2 / 2) > k_\rightarrow(\langle f \rangle)$$
$$k'_\leftarrow(\langle f \rangle, \delta) = k_\leftarrow(\langle f \rangle)\exp(\beta^2 x_2^2 \delta^2 / 2) > k_\leftarrow(\langle f \rangle). \tag{A5}$$